\documentclass[fleqn,twoside]{article}
\usepackage{espcrc2}
\usepackage{graphicx}
\usepackage{amsmath,amssymb}

\newcommand{\bl}{\protect%
\begin{list}%
{-- \hfill}{\setlength{\leftmargin .4cm} \setlength{\listparindent
0cm} \setlength{\labelwidth .2cm} \setlength{\itemsep 0cm}
\setlength{\parsep 0cm} \setlength{\topsep 0pt}\setlength{\labelsep 0.2 cm}}}
\newcommand{\el}{\protect%
\end{list}}
\newcommand{\ed}{\protect%
\end{list}}

\title{Recent development and perspectives of machines for lattice QCD}

\author{Th. Lippert\address{Department of Physics, 
        University of Wuppertal, 42097 Wuppertal, Germany}}
       
\begin{document}

\begin{abstract}
  I am going to highlight recent progress in cluster computer technology
  and to assess status and prospects of cluster computers for lattice QCD
  with respect to the development of QCDOC and apeNEXT. Taking the
  LatFor test case, I specify a 512-processor QCD-cluster better than
  $1\$/$Mflops.
\vspace{1pc}
\end{abstract}

\maketitle

\section{INTRODUCTION}

Driven by the ever increasing demand of lattice QCD for compute power,
Computer Science has become a serious activity of its own for many research
groups, with a proven record of success. A variety of ``home made'' QCD
engines is described in a long-standing series of ``machine talks'' from
early Lattice Conferences on.

In US, lattice physicists always were close to computer companies, {\em
  e.g.}\ IBM, where the {\em first-generation} GF11 project started in
1983 \cite{Beetem:1986ma,Weingarten:1990ze,Weingarten:1992nv}, or TMC
\cite{Brickner:1990qq,Aoki:1991vc}.  In Japan, cooperation of physics and
computer science began in 1978 \cite{Iwasaki:1994me} continuing with the
QCDPAX series \cite{Iwasaki:1990my}. In 1996, CPPACS, a long term leader
of the TOP500 list, was built by CCP scientists together with computer
industries \cite{Iwasaki:1998qr,top500}. Certainly, this symbiosis to a
good deal has pushed Japan to the top position in HPC
\cite{Aoki:2003gj}.

Soon fourth generation ``home-made'' systems will become operational: In
the US, computer science activities at Columbia go back to 1982
\cite{Christ:1984gp,Christ:1989bp,Christ:1990wk}.  The group has devised
QCDSP in 1999, a {\em third generation} QCD-computer
\cite{Mawhinney:2000fx}, and is about to finish the prototype of QCDOC, a
highly scalable multi-Tflops system \cite{Boyle:2002wg,Boyle:2003ue}
in collaboration with UKQCD and IBM. In Europe, INFN/Rome started with the
{\em first generation} APE in 1984 \cite{Marinari:1986si}.  In 1993, the
INFN/Pisa-Rome group has presented the {\em second generation} APE100
\cite{Marinari:1993ia,Sexton:1996xi}, followed by the {\em third
  generation} APEmille in 2001 \cite{Bartoloni:2001he}.  First CPUs for
apeNEXT, designed by the Berlin-Pisa-Rome group (DESY-INFN) for a speed of
several Tflops, are expected for autumn 2003 \cite{Ammendola:2002xt}.

A new HPC variety has entered the stage more recently
\cite{Gottlieb:2000dv,Bhanot:2001xn}.  Built from standard PC components,
cluster computers can be readily adapted to lattice QCD
\cite{Luscher:2001tx}. They strive to win the QCD-computer contest for
lowest price/performance ratios, claimed by both QCDOC and apeNEXT with a
sustained performance of 1\$/Mflops (Mflop/s) for double precision Wilson
fermion computations in 2004.

I have been asked to highlight recent progress in cluster computer technology
and to assess the opportunities of QCD-clusters and home-made systems to win the
contest.  To this end I choose the LatFor test case
\cite{Hasenbusch:2003rs} and consider two cost functions, the
price/performance ratio $R$ for investment costs and the waste heat $H$ for
cost of operation.  Based on performance results given in section
\ref{PERF}, I will specify a 512-processor QCD-cluster with $R=1\$
$/Mflops and $H\approx 0.12$W/Mflops.

In sections \ref{RISE} and \ref{OPT}, I discuss general and QCD-optimized
clusters. Recent PC hardware developments are presented in section
\ref{HARD}.  The status of QCDOC and apeNEXT is given in section
\ref{APEQCDOC}.

\section{RISE OF CLUSTER COMPUTING\label{RISE}}

\begin{figure}[!tb]
\vspace*{-.6cm}
\includegraphics[width=\columnwidth]{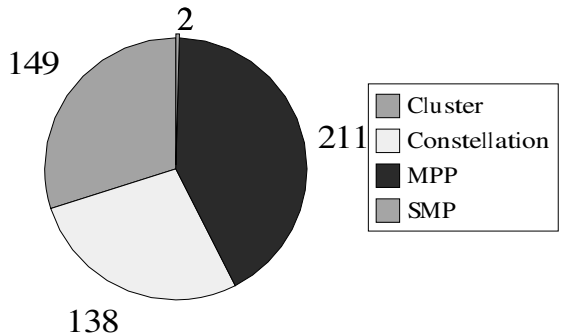}
\caption{Computer class distribution in the TOP500 list.}
\label{fig:TOP500}
\end{figure}

Table \ref{table:top500} illustrates the increasing presence of
cluster computers in the TOP500 list, which is sorted according to
Linpack benchmark results of the most powerful computing systems
worldwide \cite{top500}.
\begin{table}[!tb]
\caption{Percentage of cluster computers in the TOP500 list.}
\label{table:top500}
\renewcommand{\tabcolsep}{9.5pt} 
\renewcommand{\arraystretch}{1.2} 
\begin{tabular}{@{}cccc}
\hline
 & total & \begin{minipage}[b]{1.5cm}academic/\\[-.2cm]research\end{minipage} 
 & industry \\
\hline
Jun. 2003 & 29.7 & 20.8 & 8.9 \\
Nov. 2002 & 18.6 & 13.4 & 4.0 \\
Nov. 2001 & 8.6  & 6.2  & 1.2 \\
Nov. 2000 & 5.6  & 3.6  & 1.2 \\
Nov. 1999 & 1.4  & 1.0  & 0.0 \\
\hline
\end{tabular}
\end{table}
The TOP500 group defines a {\em cluster} as parallel computer where the
number of nodes is larger than number of processors per node.  If the
number of nodes is less than the number of processors per node
the system is termed {\em constellation}. The main fraction in the
TOP500 list consists of single node MPPs while non-clustered SMPs have nearly vanished (2\%), {\em
  cf.}\ figure~\ref{fig:TOP500}.
Among the first 100 entries, 33 clusters are found in US, 3 in France, 2 in
Sweden, and 1 in Australia, Canada, China, Germany, Russia, and UK,
respectively.

\begin{figure}[!tb]
\includegraphics[width=\columnwidth]{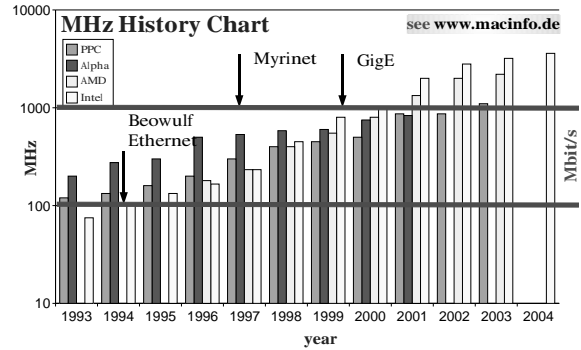}
\caption{Co-evolution of CPU clock rate and network speed.}
\label{fig:MHZ}
\end{figure}

Cluster computing started with Beowulf systems in 1994
\cite{ridge97beowulf}. But it was not before the advent of networks
with gigabit point-to-point performance like Myrinet that clusters
could become competitive. While CPU clock rates grow more or less
continuously, doubling every 21 months according to Moore's ``Law'',
performances of commodity networks tend to increase in a step-wise
fashion.  As a rule of thumb, many HPC applications ask for a network
speed of 1 Gbit/s per 1 GHz clock speed of a node.  Fig.~\ref{fig:MHZ}
demonstrates that this matching point was reached around 1999
\cite{MACINFO}.
The breakthrough of clusters occurred when PC prices went down and low
latency switches, high level message passing standards like MPI and reliable
communication software became available.

Clusters bear several advantages: they are built from cost-effective
components, their modularity allows flexible hardware upgrades, they
benefit from (OpenSource) software standards and they can be optimized for
many specific applications.

\section{QCD-OPTIMIZED HPC-CLUSTERS\label{OPT}}

\begin{figure}[!tb]
\centerline{\includegraphics[width=.7\columnwidth]{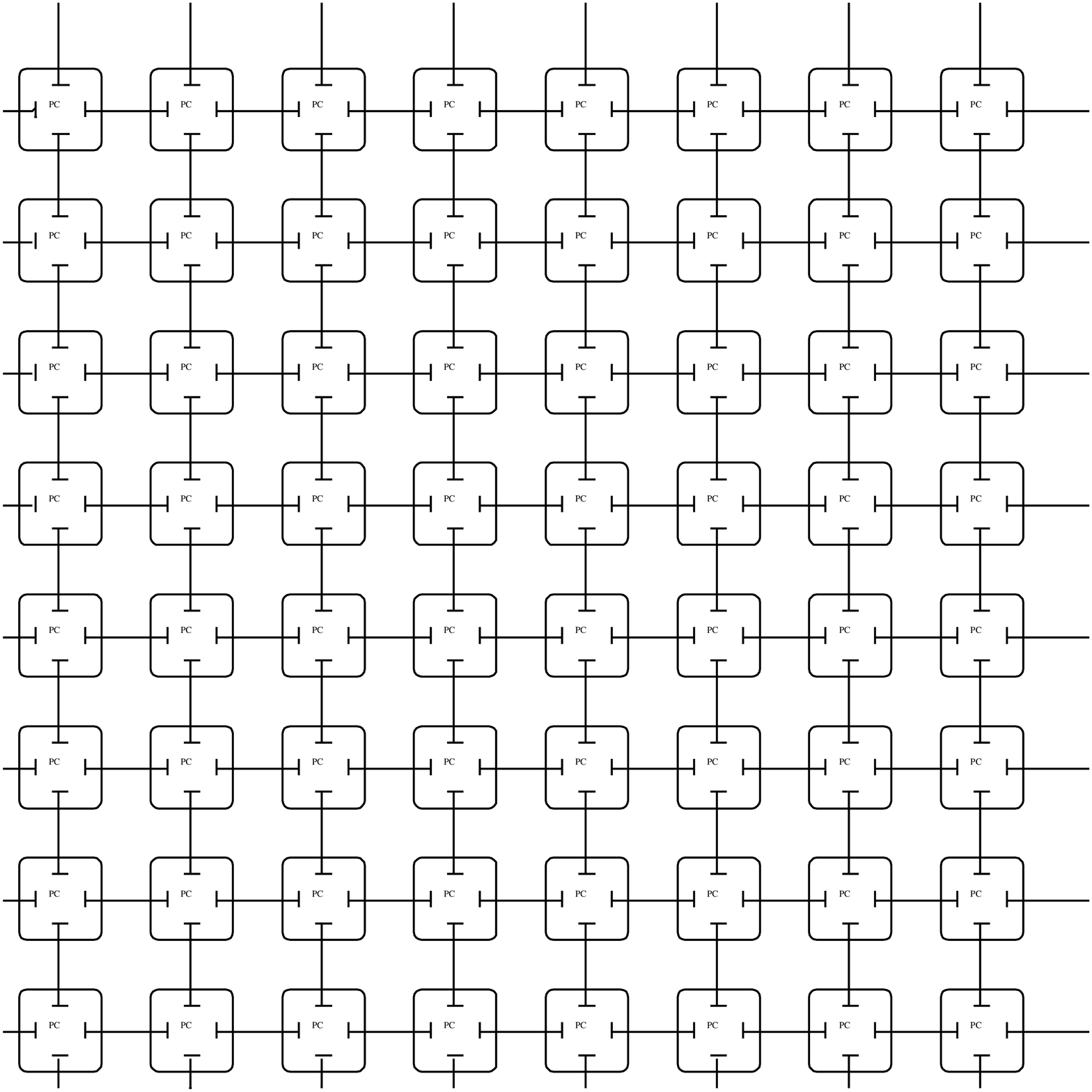}}
\vspace*{-.8cm}
\caption{GigE mesh of the Budapest PMSv.3 cluster (from 
  \cite{Fodor:2002zi}).}
\label{FODORMESH}
\end{figure}

QCD-Cluster computing was pioneered by Gott\-lieb in 1998.  He built the
``Candycane'' Beowulf from 32 350 MHz PentiumII PCs.  Other early systems
followed soon, see table~\ref{table:early}.
\begin{table*}[!tb]
\caption{``Early'' QCD-clusters.}
\label{table:early}
\renewcommand{\tabcolsep}{2pt} 
\renewcommand{\arraystretch}{1.05} 
\begin{tabular*}{\textwidth}{@{}cccccccc@{}}
\hline
site & name & year & \# of procs & CPU type & clock [MHz] & net & \\
\hline
Indiana State &Candycane &  1998 & 32 & PII & 350 & fast ethernet &
 \cite{GottliebCANDY} \\
 E\"otv\"os/Budapest & PMS & 1998/1999 & 32/64 & K6-2 & 450 & ISA 2D & 
 \cite{Csikor:1999vz} \\
Wuppertal & ALiCE &  1999 & 128 & Alpha 21264 & 616 & Myrinet &
 \cite{PIK:2002} \\
Jlab & Calico &  2000 & 16 + 18  & Alpha 21264 & 667 & Ethernet & \cite{CALICO}\\
 Adelaide & ORION &2000 & 40 $\times 4$ & SUN E420R& & Myrinet2000 &
 \cite{ADELAIDE} \\
FNAL & QCD80 & 2000 & 80 & PIII dual & 700   & Myrinet2000 & \cite{QCD80}\\
 Zhongshan/Guangzhou &       &2000 & 10 & PIII dual & 500  & fast ethernet & \cite{Luo:2002kr}\\
\hline
\end{tabular*}
\end{table*}
Since 2002, quite a few QCD-clusters  have been installed.
Still, the number of systems and their individual sizes are small
compared to the general purpose clusters of the TOP500 list:  \vspace{3pt}
\bl
\item[$\bullet$]
Bielefeld \cite{Bielefeld} (2003)
\bl
\item[-] 16 dual XEON, 2.4 GHz 
\item[-] with switched GigE
\item[-] 16 dual Athlon MP 1800
\item[-] with Myrinet2000
\el
\item[$\bullet$]
Bern \cite{BERN} (2003)
\bl
\item[-] 32 dual XEON, 2.4 GHz 
\item[-] Intel E7500 chipset
\item[-] DDR RAM
\item[-] Myrinet2000, 2 $\times$ 190 MB/s bi-dir bw
\el
\item[$\bullet$]
DESY \cite{DESY} (2002)
\bl
\item[-] 16+16 dual XEON, 1.7/2.0 GHz (Hamburg)
\item[-] 16 dual XEON, 1.7 GHz (Zeuthen)
\item[-] Supermicro P4DC6
\item[-] 1 GB RDRAM per node
\item[-] Myrinet2000 
\el
\item[$\bullet$]
FNAL \cite{FNALNEW} (2002)
\bl
\item[-] 128 dual XEON, 2.4 GHz 
\item[-] 48 dual XEON, 2.4 GHz
\item[-] Supermicro P4DPR-6GM+
\item[-] Intel E7500 chipset
\item[-] 128 + 48 GB DDR RAM
\item[-] Myrinet2000, 2 $\times$ 135 MB/s bi-dir bw
\item[-] GigE mesh on 16 nodes
\el
\item[$\bullet$]
Jlab \cite{JLABNEW} (2002)
\bl
\item[-] 128 single XEON, 2.0 GHz 
\item[-] Intel E7500 chipset
\item[-] 65 GB DDR RAM
\item[-] Myrinet2000
\el
\item[$\bullet$]
Seoul  (2003)
\bl
\item[-] 30 P4, 2.4 GHz 
\item[-] 16 GB DDR RAM
\item[-] Fast ethernet
\el
\item[$\bullet$]
Taipei \cite{Chiu:2002bi} (2003)
\bl
\item[-] 30 P4, 1.6/2.0 GHz, Farm 
\item[-] RDRAM tuned for overlap simulations
\el
\item[$\bullet$]
Tsukuba (CCP) \cite{TSUKUBSCLUSTER} (2003)
\bl
\item[-] 16 dual XEON, 2.8 GHz 
\item[-] 64 GB DDR RAM
\item[-] Myrinet2000
\el
\el
\vspace{3pt}
In early 2002, the Budapest group has carried out first runs with the {\em
  Poor Man's Supercomputer v.3} (PMS v.3). The {\em Budapest Architecture}
was the first large cluster system to use a Gigabit ethernet mesh as
connectivity: 
\vspace{3pt} 
\bl
\item[$\bullet$]
Budapest \cite{Fodor:2002zi} (1/2002)
\bl
\item[-] PMSv.3 
\item[-] 128 P4, 1.7 GHz
\item[-] Intel GBD MB
\item[-] 512 MB/node RDRAM
\item[-] 4 $\times$ SMC 9452 Gbit-NIC
\item[-] PCI 32bit/33 MHz
\el
\el
\vspace{3pt}
The Budapest Architecture can be reconfigured to smaller partitions by
re-wiring of the GigE-mesh, see figure \ref{FODORMESH}.
\begin{figure}[!tb]
\includegraphics[width=\columnwidth]{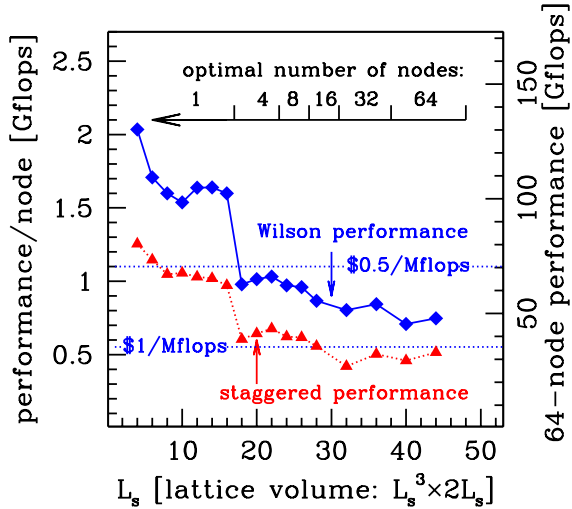}
\vspace*{-.6cm}
\caption{Performances of Wilson and Staggered fermion matrix inversion on
  the PMSv.3 (from \cite{Fodor:2002zi}).}
\label{FODORWILSON}
\end{figure}
PMSv.3 achieves a price performance ratio of less than $1 \$/$Mflops
sustained for {\em single precision} Wilson fermion matrix inversions. This
ratio has been assessed in 1/2002 by pricing data quoted at {\tt
  www.pricewatch.com} adding 10 \%.  Fig.\ \ref{FODORWILSON} shows node
performances for an optimal number of nodes constrained by the available
memory on PMSv.3.  The MILC HMC code was optimized by SSE constructs for
time-critical code blocks\footnote{The use of the multi media extension
  (MMX) for AMD K6-2 has been suggested in 1999
  \cite{Csikor:1999vz} for PMSv.1 and was subsequently used in finite
  density QCD computations \cite{Fodor:2001pe}.  At the same time, M.\ 
  L\"uscher has presented fast SSE coding on Intel platforms
  \cite{Luscher:2001tx}.}.

PMSv.3 has demonstrated that sophisticated networks can be avoided on
streamlined QCD-clusters, which otherwise eat up a substantial part of the
available budget. 

The Budapest Architecture is a model for QCD-clusters ({\em i}) providing
high single node performances, ({\em ii}) delivering sufficient network
performance at low costs, ({\em iii}) and being scalable.

\begin{table*}[!htb]
\begin{center}
\includegraphics[width=.7\textwidth]{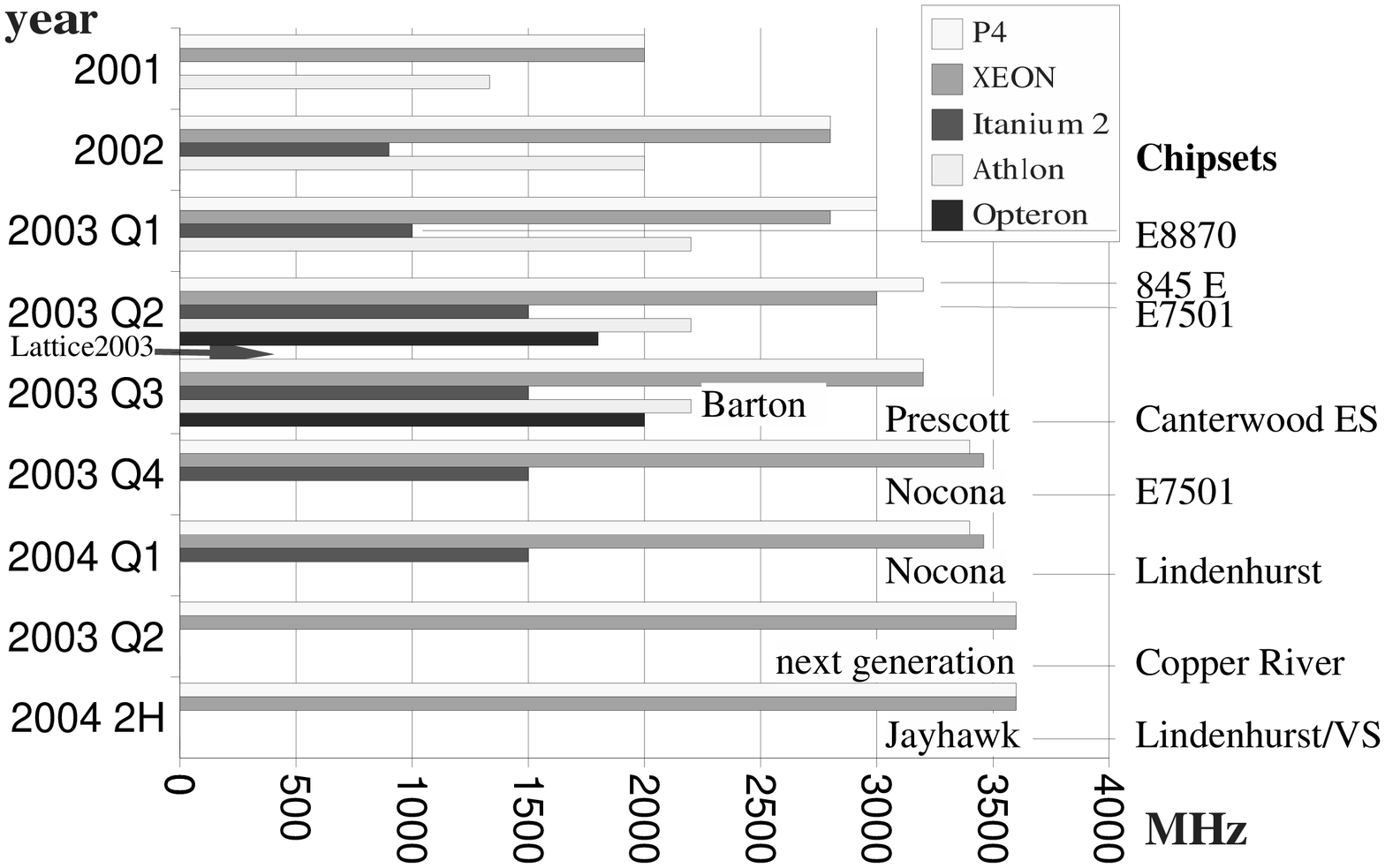}
\end{center}
Figure 5. Clock frequency road-map (status Q2 2003).
\end{table*}

\stepcounter{figure}

\section{CLUSTER HARDWARE TRENDS\label{HARD}}

The efficiency of parallel computers is determined both through the local
efficiency of the compute nodes and the performance of the communication
network. In particular, the speed of the network interface, {\em i.e.}\ the
speed of the PCI sub-system, is a key parameter to benchmark present
commodity hardware.

\subsection{CPUs}

Let's concentrate on PC processors that are currently relevant for cluster
computing: in Q2 2003, clock frequencies of Intel P4 and XEON CPUs have
reached more than 3 GHz; AMD Athlon and Opteron CPUs have touched the 2 GHz
threshold\footnote{The number tags of AMD Athlons mimick the
  performance-equivalent clock rate of Intel chips.}; The Athlon64 CPU
appeared in Q3 2003. While it's safe to say that the development of CPU
clock speeds will follow Moore's ``Law'', it is of course difficult to
predict the detailed evolution of CPUs and chipsets, even for the near
future.  In Fig.\ 5, I have tried to collect the information made public by
Intel and AMD.  According to these numbers, Intel P4 and XEON processors
will approach 3.4 GHz near the end of 2003 while CPU speeds of more than
3.6 GHz cannot be envisaged before Q2 2004.  The 1.5 GHz Itanium2 chip
appears to be with us for quite a while. A successor to the Pentium 4,
called ``next generation'' processor, might be expected in the second half
of 2004.  Further details on Intel and Opteron processors cannot be given
here.

\subsection{Memory and front side bus\label{MFSB}}

QCD computations are largely determined by the memory-to-cache data rate
available on the given chipset. A key figure is the frequency of the
so-called {\rm front side bus} (FSB). The FSB connects the processor to the
north-bridge, the memory controler hub (MCH). The memory frequency itself
must match the FSB frequency for maximal bandwidths.  To give
an example, the 800 MHz FSB requires 400 MHz dual channel DDR RAM (PC3200)
to be fully saturated.

Let's clarify the nomenclature: The acronym DDR stands for ``double
data rate'' exploiting both the rising and falling flanks of the
signal unlike standard SDRAM.  Such memory type in principle delivers
a data rate $D=8\times 2 \times f$ B/s.  As a next step, the {\em
  dual channel} memory controlling technology has been introduced which
allows to double $D$ once more by means of logical words of length 144
bits ($2\times (64 + 8)$) that are split over two memory banks. In
other words, a dual channel twin module mimicks an effective frequency
of $2f$ MHz. With respect to the above example, $f=400$ MHz is the
effective frequency of a twin module with $D=6.4$ GB/s.

A further increase in DDR memory frequency is expected for Q1 2004
with the appearance of 667 MHz DDRII dual channel SDRAM. Recently,
RAMBUS announced XDR DRAM running at a speed from 3.2 to 6.4 GHz with
$D=6.4$ and 12.8 GB/s per channel \cite{XDRRAM}. But keep in mind that
RAMBUS memory tends to be nearly twice as expensive as equivalent DDR memory.

\begin{figure}[!htb]
\centerline{\includegraphics[width=.8\columnwidth]{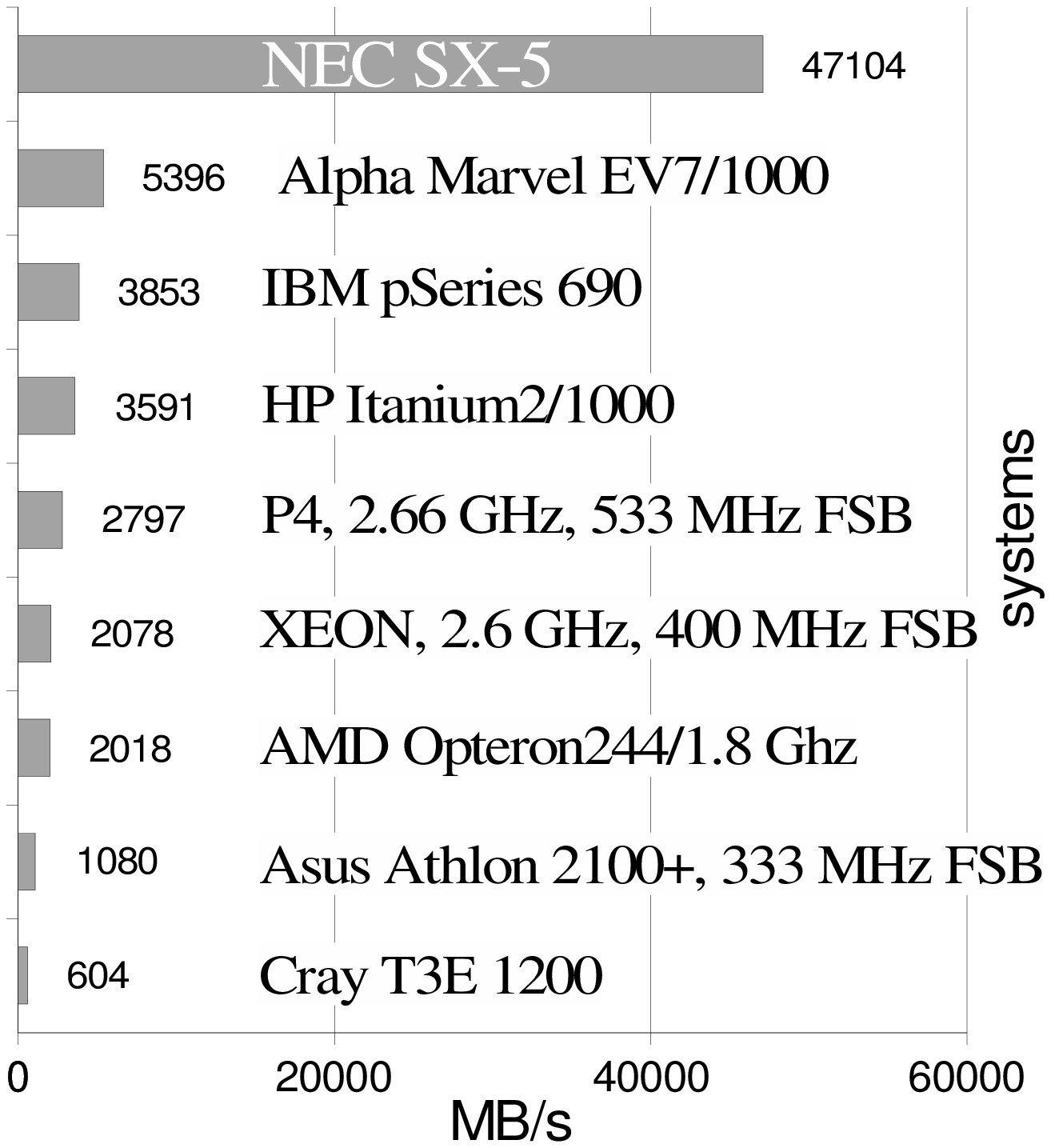}}
\vspace*{-.6cm}
\caption{Comparison of STREAM ``Triad'' benchmark results on a variety of
  computer platforms (taken from \cite{STREAMBENCH}).\vspace*{-.6cm}}
\label{FIG:STREAMS}
\end{figure}

The STREAM benchmark is a reliable estimate for the actual data rate that
can be achieved on a given system. Fig.\ \ref{FIG:STREAMS} shows results of
STREAM for a variety of platforms \cite{STREAM}.
In terms of the maximal bandwidth, STREAM gives about 87\% on a 2.66 GHz P4
platform equipped with 200 MHz DDR RAM on a 533 MHz FSB, for example.  The
STREAM benchmark is clearly dominated by the NEC SX-5 vector system.

At this stage let me note that the so-called ``machine
balance''\footnote{$B$ should better be called machine imbalance.} $B$,
{\em i.e.}\ the ratio of Mflops vs.\ the memory accesses in Mwords/s,
has increased for PC hardware in recent years.  While Intel 486 boards
had $B$ close to 1---a value today maintained on vector systems like
the SX-5 only---Pentium P4 boards show $B= 10\dots 20$.  Consequently,
a 1.7 GHz P4 CPU saturates 2 channel RAMBUS 400 MHz RAM with a maximal
data rate of 1.6 GB/s for SSE-boosted Wilson fermion codes.  It
currently appears to be less cost efficient to choose CPUs with
highest frequency.

$B$ is even more unfavorable for XEON dual processor systems as the
FSB capacity is shared among the processors.  Given a maximal FSB
frequency of 533 MHz (Q3 2003) $B$ is nearly 3 times larger than for
fastest Pentium P4 boards with 800 MHz FSB, which were available
already in Q2 2003.  800 MHz boards for the XEON processor will not
become available before Q2 2004, still the difference to P4 will be a
factor of two\footnote{The catch is the weak PCI bus of PC boards.}.

AMD currently supports a FSB frequency of 400 MHz for the Athlon processor
while the memory connection to the AMD Opteron processor is enabled through
an internal memory controler with direct memory access. The advantage is
that $B$ is constant for single, dual or quad Opteron systems.  In
other words, Opteron is scalable.

\subsection{PCI\label{PCI}}

PCI is a hardware standard to connect PCs with external devices.  The
speed of PCI is the bottleneck dominating the performance of the
interconnectivity of cluster computers.  We have witnessed several
improvement steps since 1993 through which PCI evolved from a
32bit/33MHz bus to 64bit/133MHz PCI-X in 1999.  However, one should be
aware that 64bit PCI bus widths are not supported on standard PC
boards.  Clearly, the theoretical bi-directional Gigabit-Ethernet
performance of 2 Gbit/s cannot be served adequately by a 32bit/33MHz
PCI bus. Thus, already for Gigabit-Ethernet we encounter a 2:1 PCI-bus
over-booking on standard PC boards. Myrinet2000 with a bi-directional
bandwidth of 4 Gbit requires at least a 64bit/66MHz PCI
bus\footnote{To my knowledge there are is only one genuine P4 board
  with a FSB above 500 MHz supporting PCI-X, while many boards for
  dual XEON, Athlon and Opteron processors meanwhile are equipped with
  PCI-X. The Tyan 2726 XEON board even supports 4 on-board GigE slots on 2
  PCI-X channels \cite{TYAN2726}.}.

A new standard, PCI-Express, is about to enter the market early in
2004.  PCI-Express is a fundamental re-design as compared to PCI-X.
Instead of a parallel 64bit bus, PCI-Express is based on a serial bus
with several channels (lanes). The performance per lane will be 
2.5 Gbit/s, up to 32 lanes are possible. With PCI-Express cluster
nets will enter the O(100) Gbit/s era \cite{PCIEXPRESS}.

It is well known that early P4 and XEON board chipsets were delivering much
less PCI-bandwidth than promised by specifications. On today's Intel E750x
and Serverworks GC chipsets such performance degradations 
have been overcome \cite{LINDAHL}.

\subsection{Network technology}

\begin{figure}[!tb]
\includegraphics[width=\columnwidth]{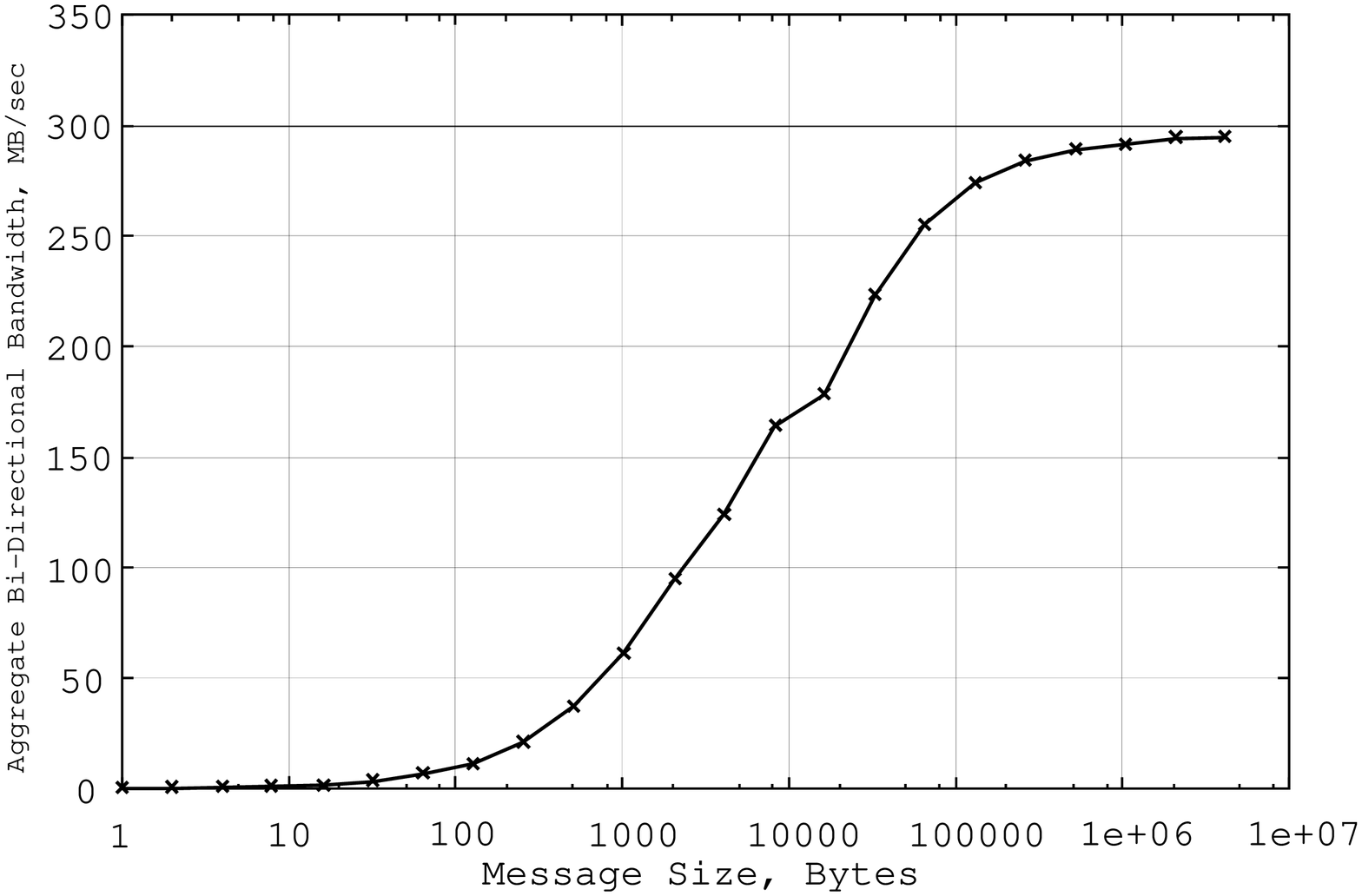}
\vspace*{-.6cm}
\caption{Myrinet aggregate bi-directional bandwidth on a XEON system
  for the PALLAS MPI-benchmark (taken from \cite{FNALNEW}).}
\label{FIG:MYRI}
\end{figure}

Cluster pioneer Myricom presented ``Myrinet'' with  2 Gbit/s
bi-directional bandwidth already in 1997 and has evolved the product to
Myrinet2000 with  bi-directional bandwidth of 4 Gbit/s. Fig.\ 
\ref{FIG:MYRI} shows the aggregate MPI-bandwidth as a function of the
message length, using the genuine Myrinet communication driver GM 1.5.3 on
the FNAL systems, {\em cf.}\ section \ref{OPT}.
As just announced, the maximal bi-directional bandwidth can reach 950
MB/s (two channels).  The latency for the PALLAS
MPI-benchmark, {\em i.e.}\ the half of the zero-message length
round-trip time, is 5 $\mu$s \cite{MYRI}.  Switches are available for
up to 128 ports in a single cabinet. They can be combined to
multi-stage crossbars with thousands of ports. The switch latency lies
in the range of O(100) ns per stage.

\begin{figure}[!tb]
\includegraphics[width=\columnwidth]{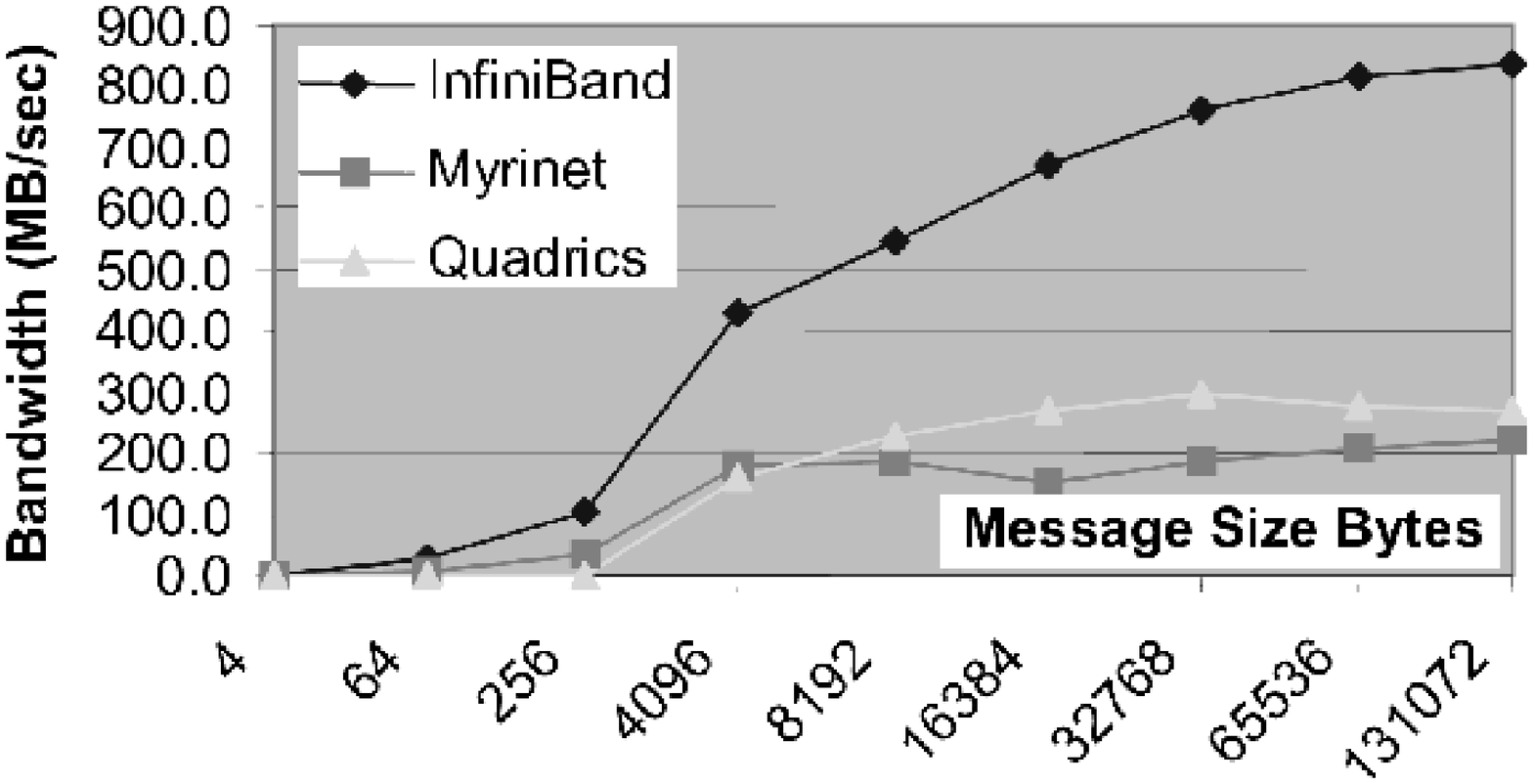}
\vspace*{-.6cm}
\caption{Comparison of aggregate bi-directional bandwidths 
  (PALLAS MPI-benchmark) for Infiniband, Myrinet2000 and QsNET (Quadrics)
  (taken from \cite{MELLA}).}
\label{FIG:MELLANOX}
\end{figure}

\begin{figure}[!tb]
\vspace*{-.6cm}
\includegraphics[width=\columnwidth]{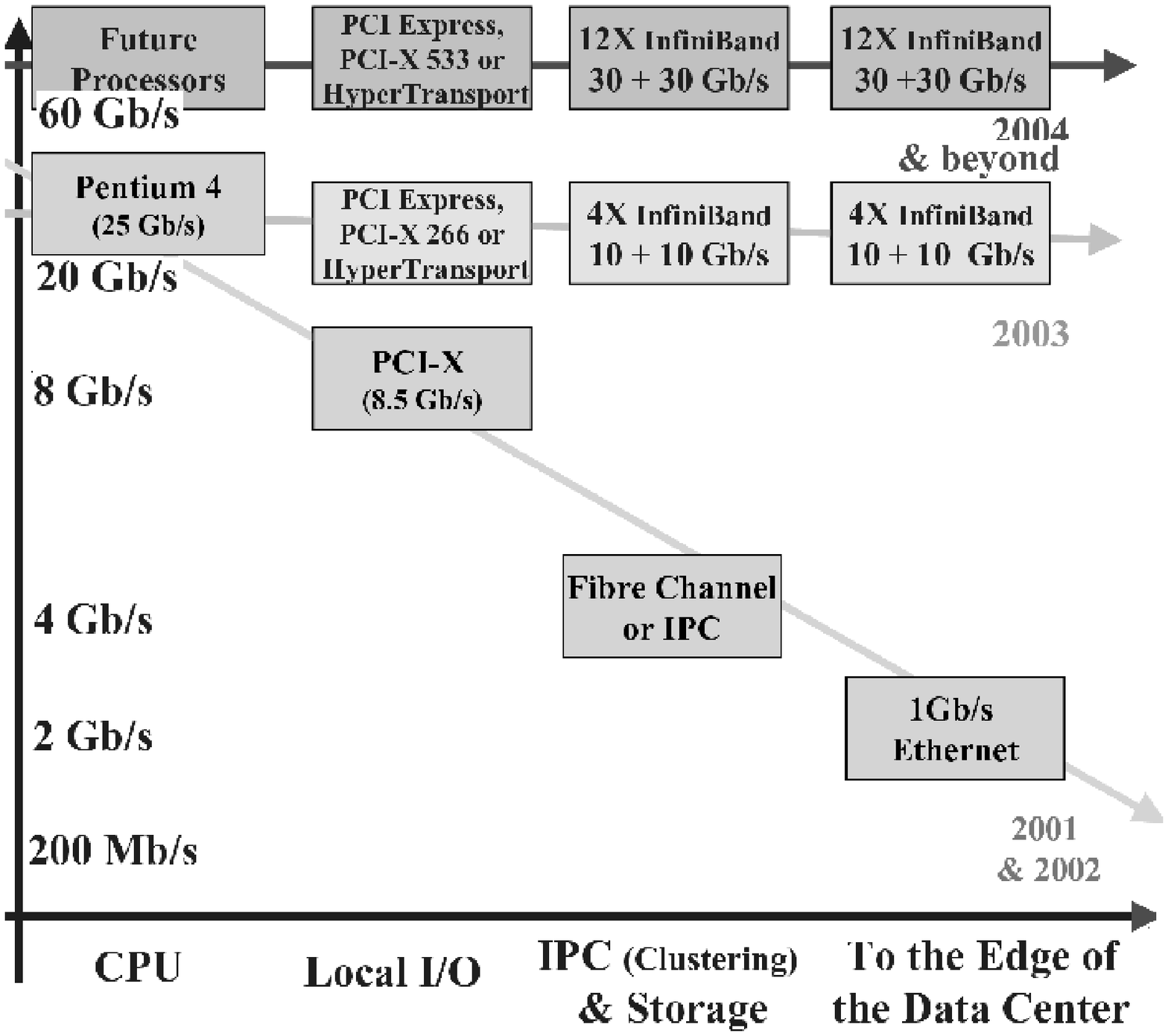}
\vspace*{-.6cm}
\caption{Infiniband/PCI-Express road-map for given
  ``distances'' (adapted from  \cite{MELLA}).}
\label{FIG:MELLANOX2}
\end{figure}

Another major advance in cluster network performance has been
achieved with the novel Infiniband standard. Infiniband is designed
for a bandwidth of 10 Gbit/s. First performance measurements can be
found in Ref.~\cite{MELLA} (Fig.\ \ref{FIG:MELLANOX}).  The latency
for zero-message length is supposed to be about 7 $\mu$s.  The
Infiniband road-map is shown in Fig.\ \ref{FIG:MELLANOX2}.
With PCI-Express expected for early 2004, Infiniband networks will deliver
up to 20 Gbit/s bi-directional bandwidth.  Currently 96-port switches are
available that can be combined to a larger multi-stage crossbar. The
additional latency per switch stage is reported to be about 200 ns.

While the performances of Myrinet, QsNET and Infiniband are
impressive, the costs are substantial. A Myrinet2000 interface card
costs \$1000 and a switch port about \$400 on average.  Mellanox
Infiniband lies in the same price range (Q3 2003). With about \$1400
per node, networking costs surpass the costs for the compute nodes. At
this stage, these sophisticated networks appear to be reserved for
high-end general purpose cluster systems.

In order to provide cheaper and faster communication the FNAL group has
constructed own Gigabit-Ethernet network cards based on FPGAs
\cite{DONHOLM}. As these cards support up to 8 ports one can arrange the
processing elements in form of a hypercube.  Currently, the card is
designed for 32bit/33MHz PCI, a PCI-X version is planned.  Still, the
costs exceed \$500 per card.

On the other hand, standard GigE PCI cards cost about \$40, while dual and
quad cards amount to \$150 and \$400, respectively. I already mentioned a
system supporting up to four  GigE ports on board, hence PCI cards
aren't required at all in case of a Budapest Architecture.

\begin{figure}[!tb]
\includegraphics[width=\columnwidth]{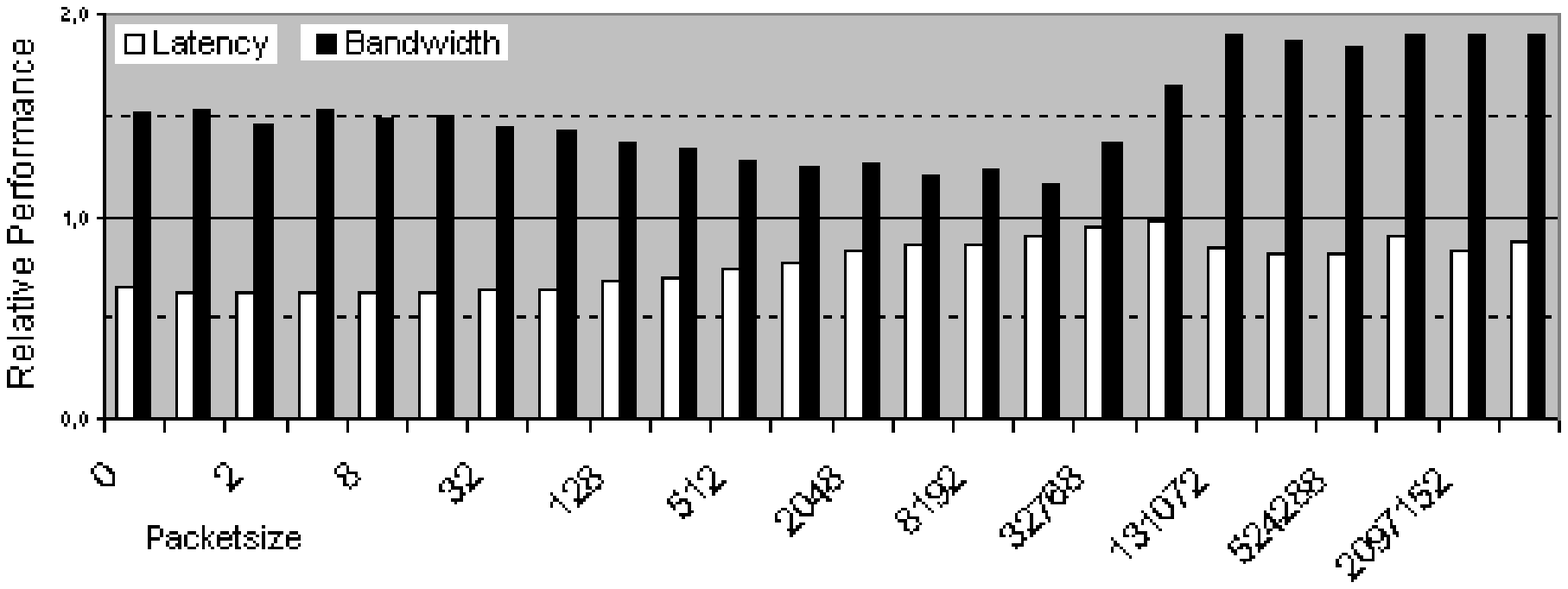}
\vspace*{-.6cm}
\caption{ParaStation TCP/IP bandwidth improvement vs.\
  standard TCP/IP under Linux on a 2.6 GHz, 400 MHz FSB XEON system.}
\label{FIG:PARA}
\end{figure}

How large a bandwidth can we squeeze out of a point-to-point GigE
connection? The answer is largely dependent on the TCP/IP driver used.
Standard drivers allow for somewhat more than 100 MB/s bi-directional
bandwidth. Communication optimized drivers like ParaStation \cite{PARTEC}
can provide a much larger bandwidth. On a XEON system (64bit/66MHz PCI),
ParaStation TCP/IP reaches up to 200 MB/s bi-directional bandwidth, see
Fig.\ \ref{FIG:PARA}.
Of course, one would wish to achieve an aggregate bandwidth of 800
MB/s on a GigE mesh. This requires, first of all, systems with PCI-X,
and in order to achieve maximal bandwidth, the communication software
has to drive four network cards simultaneously.

While meshes or grids are scalable with respect to nearest-neighbor
computations, more complicated QCD applications require a switched
network.  Quite recently, level 3 enabled {\em routed} GigE switches
appeared with O(500) ports like the Myrinet GigE switch \cite{MYRI},
the CISCO Catalyst 6500 \cite{EANTC} or the Force10 E series
\cite{FORCE10}. The costs per port came down within the last half year
(about \$300 per port in Q3 2003).  Certainly, the bi-directional
bandwidth will not exceed 200 MB/s for switched GigE connections. In fact,
most switches are overbooked. The {\em additional} latency of the
Myrinet switch is 3.5 $\mu$s, CISCO Catalyst 6500 adds between 12 and
16 $\mu$s while the FORCE10 switch is reported to give 23~$\mu$s.

Table \ref{TAB:NETS} presents throughputs and
latencies of the various PCI-based cluster connectivities.
\begin{table}[!b]
\caption{Network characteristics.}
\label{TAB:NETS}
\renewcommand{\tabcolsep}{1pt} 
\renewcommand{\arraystretch}{1.2} 
\begin{tabular*}{\columnwidth}{@{}ccccc}
\hline
Net & bw bi-dir & latency & per stage &  \\
\hline
Infiniband & 20 Gbit/s   & 7 $\mu$s & 200 ns   & \cite{MELLA} \\
QsNET     & 5.44 Gbit/s &  2 $\mu$s &          & \cite{QUADRICS} \\
Myrinet    & 4 Gbit/s    & 5 $\mu$s & 200 ns & \cite{MYRI}\\[-.2cm]
(2003)    & 8 Gbit/s     \\
\hline
GigE      & 2 Gbit/s     & 27 $\mu$s & 12-23 $\mu$s &  \\[-.2cm]
(ParaStation)      &     & 12 $\mu$s &  & \cite{PARTEC} \\[-.2cm]
(JLAB)             &     & 12 $\mu$s &  & \cite{WATSONPRIVATE} \\
\hline
\end{tabular*}
\end{table}

A comment: clusters with  hybrid networks, {\em i.e.}\ merging a
mesh with a switched system, appear to be  quite an effective solution for
non-nearest-neighbor QCD computations.  In that case, nearest-neighbor
communication can be routed over the mesh, non-nearest-neighbor
communication tasks are routed through the switch.

\subsection{Middleware}

One should not forget stability and administration of clusters.
These issues, which become crucial on large systems, are the domain of {\em
  cluster middleware} like SCore \cite{SCORE} or Para\-Station
\cite{PARTEC}.  Besides error correction, package-loss-safety---though
expensive to realize---is required to achieve long term stability.
Furthermore, large systems need automatized administration tools which can
take care for safe job termination and system supervision.  To give an
example, the ParaSation middleware is based on a virtual machine/partition
concept, that can prevent local instabilities from 
spreading.  In this manner, we enjoy stable uptime periods of several months on
the Wuppertal ALiCE cluster.

\section{PERFORMANCE AND SCALING OF QCD CODES\label{PERF}}

\subsection{Single node performance}

\begin{figure}[!htb]
\centerline{\includegraphics[width=.8\columnwidth]{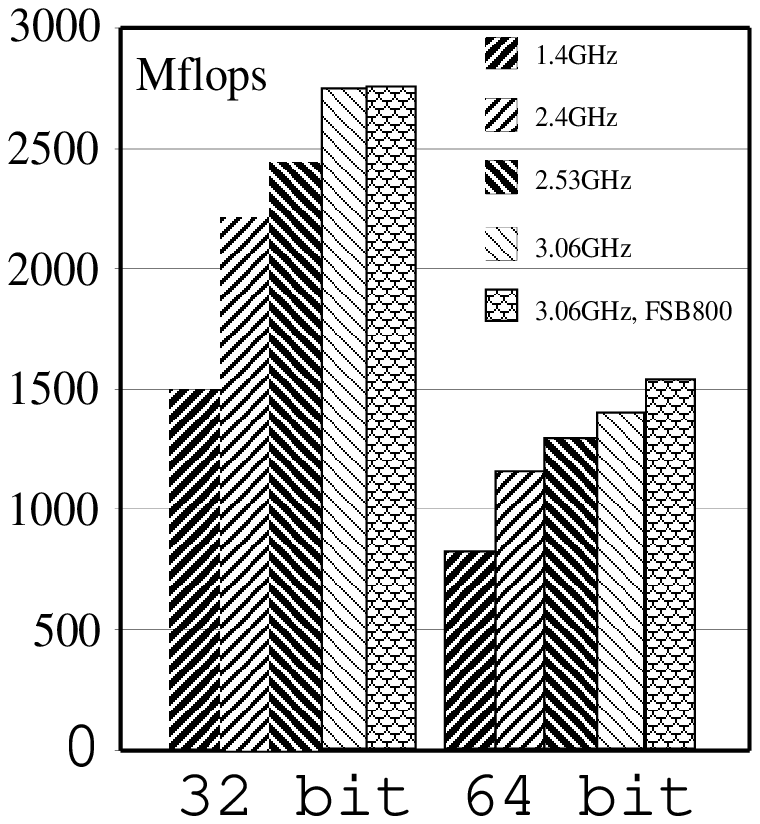}}
\vspace*{-.6cm}
\caption{Performance of the Wilson-Dirac 
  multiplication as function of the CPU clock at fixed lattice size
  $16^3\times 16$ (adapted from \cite{Gellrich:2003ps} and \cite{WEGNER}).}
\label{FIG:HASENBUSCH}
\end{figure}

The acceleration of QCD codes on single CPUs is of primary concern in order
to achieve a high parallel performance. We can benefit from Moore's ``Law'':
Fig.~\ref{FIG:HASENBUSCH} demonstrates the performance improvements gained
through increases of processor frequencies for the matrix-vector
multiplication on a $16^3\times 16$ lattice with the Wilson-Dirac operator
using 1 processor per node (code by M.\ Hasenbusch)
\cite{Hasenbusch:2003rs,Gellrich:2003ps,WEGNER}.
Performance critical parts of the Wilson-Dirac kernel are accelerated by
SSE and SSE2 (streaming SIMD extension) constructs as described in
Ref.~\cite{Luscher:2001tx}.

\begin{figure}[!tb]
\centerline{\includegraphics[width=\columnwidth]{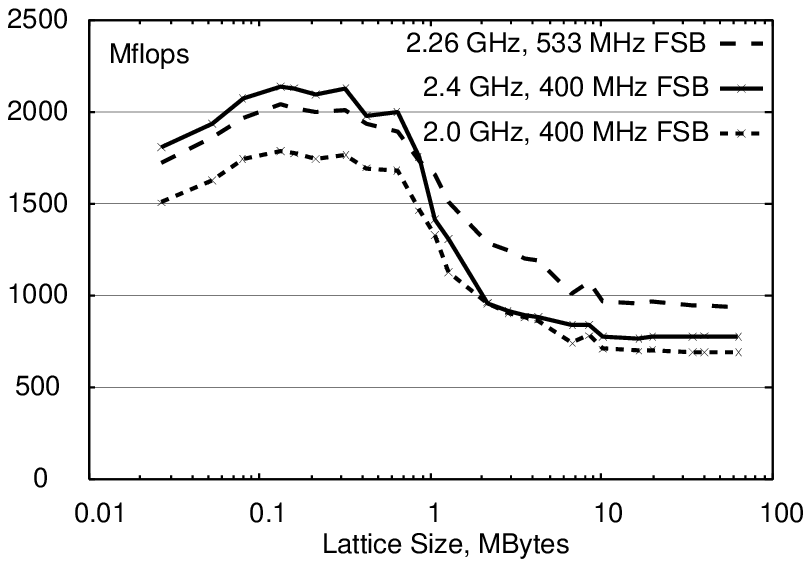}}
\vspace*{-.6cm}
\caption{Performance of staggered fermion
  multiplication as function of the CPU clock (taken
  from \cite{Singh:2003qh}).}
\label{FIG:HOLMGREN}
\end{figure}
\begin{figure}[!tb]
\vspace*{-.6cm}
\centerline{\includegraphics[width=\columnwidth]{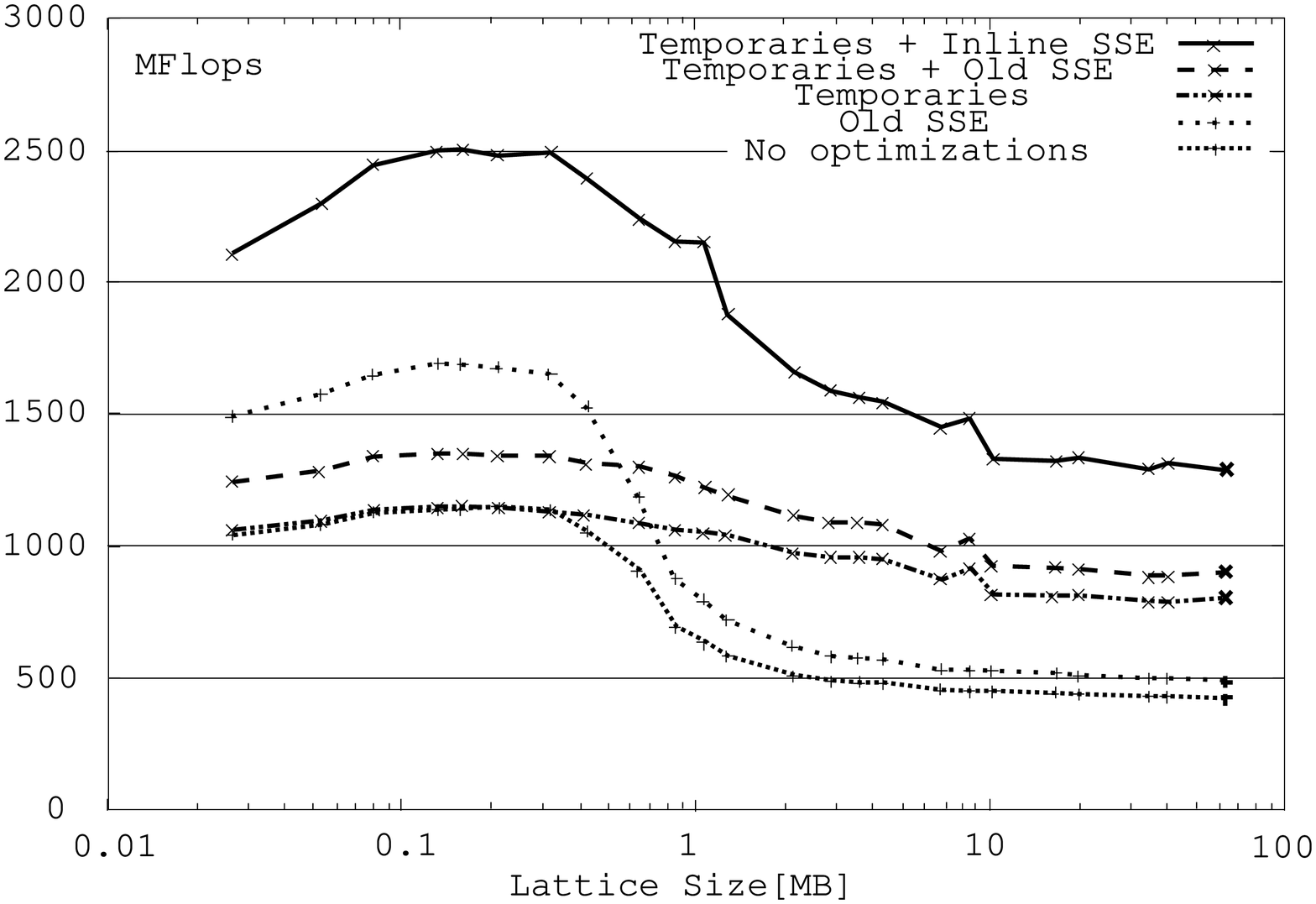}}
\vspace*{-.6cm}
\caption{Demonstration of successive performance optimization 
on the Pentium 4 with 800 MHz FSB for staggered fermions 
\cite{GOTTLIEBPRIVATE}.}
\label{FIG:GOTTLIEB}
\end{figure}

Fig.~\ref{FIG:HOLMGREN} shows the dependency of the MILC staggered fermion
code (32 bit) performance on the CPU clock frequency.  
Successive performance improvements are illustrated in
Fig.~\ref{FIG:GOTTLIEB} \cite{GOTTLIEBPRIVATE}.

\begin{table*}[!htb]
\begin{center}
\includegraphics[width=.8\textwidth]{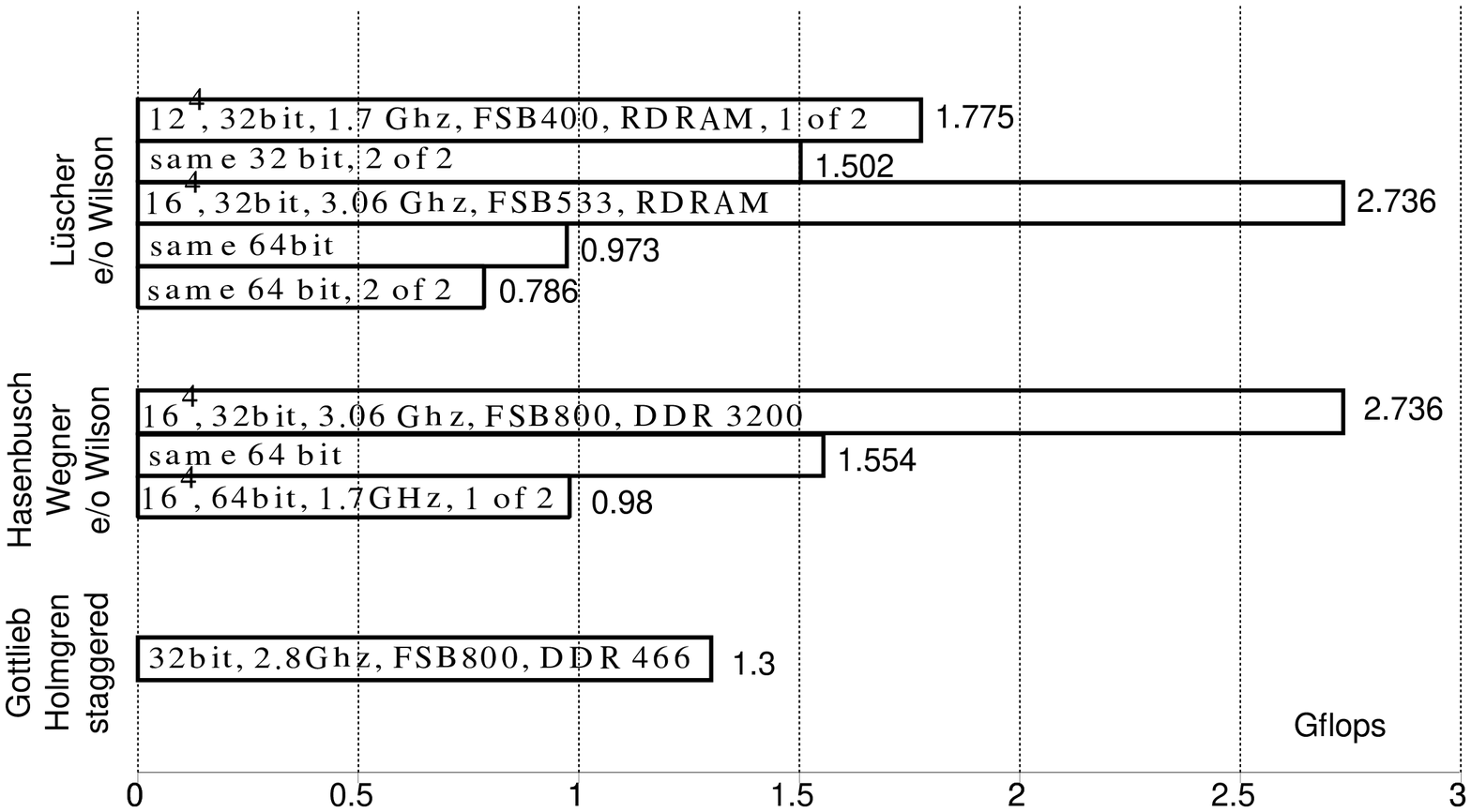}
\end{center}
Fig.\ 14.  Single processor performance records (I thank 
S.\ Gottlieb, M.\ Hasenbusch, D.\ Holmgren, M.\ L\"uscher, and P.\
Wegner for their contributions.).
\end{table*}

\stepcounter{figure}

Note that XEON (1 processor of 2, 533 MHz FSB) and P4 (1 processor of 1,
800 MHz FSB) performances differ by a factor slightly less than the FSB
frequency ratio. As to the dual node efficiency one encounters typical gain
factors---{\em i.e.}\ the gain seen when switching on the second processor and
running programs in parallel---of 1.2 to 1.4 on XEON systems with DDR RAM and
1.6 for an early dual P4 RAMBUS platform.  The small factor in case of XEON
has been anticipated below (section~\ref{MFSB}) as dual XEON processors
share the FSB. The records in local performances as of Q3 2003 are
collected in Fig.~14.

\subsection{Parallel efficiency\label{PA}}

The performance per processor will decrease for parallel operation.
On the PMSv.3 with GigE connectivity, the degradation is about a
factor of 2 for both staggered and Wilson fermions, {\em cf.}\ Fig.\ 
\ref{FODORWILSON}. The parallel efficiency, determined keeping the
local lattice size constant for single and parallel mode, is listed
in table~\ref{TAB:PARALLEL}.\footnote{The i860 chipset shows a smaller
  parallel efficiency due to the defective PCI implementation,
  mentioned earlier (section \ref{PCI}).}
\begin{table}[!b]
\caption{Parallel performances and scaled efficiency
\cite{Gellrich:2003ps}.}
\label{TAB:PARALLEL}
\renewcommand{\tabcolsep}{0pt} 
\renewcommand{\arraystretch}{1.1} 
\begin{tabular*}{\columnwidth}{@{}lccc}
\hline
system & single proc.\ & parallel      & efficiency \\[-.19cm]
       & \small [Mflops]    & \small [Mflops/proc] &  \\
\hline
Myrinet  &  579 & 307 & 0.53 \\[-.15cm]
i860, SSE&  &  & \\

Myrinet GM &  631 & 432 & 0.68 \\[-.15cm]
E7500, SSE&  &  & \\

Myrinet     \\[-.15cm]
Parastation & 675 & 446 & 0.66\\[-.15cm]
E7500, SSE  \\

Myrinet     \\[-.15cm]
Parastation &  406 & 368 & 0.91\\[-.15cm]
E7500,non-SSE \\[-.15cm] 
non-blocking \\

Gigabit \\[-.15cm]  
Ethernet & 390 & 228 & 0.58\\[-.15cm] 
non-SSE  \\
Infiniband & 370 & 297 & 0.80\\[-.15cm]
non-SSE  \\
\hline
\end{tabular*}
\end{table}
On Myrinet clusters, typically more than 65\%
efficiency are achieved for both SSE and non-SSE coding. On the
Wuppertal XEON cluster PAN (2.6 GHz, Myrinet) non-blocking
communication is enabled under MPI by virtue of ParaStation. Hence,
the communication can be hidden behind computation leading to an
efficiency of 0.91.

\begin{figure}[!htb]
\centerline{\includegraphics[width=\columnwidth]{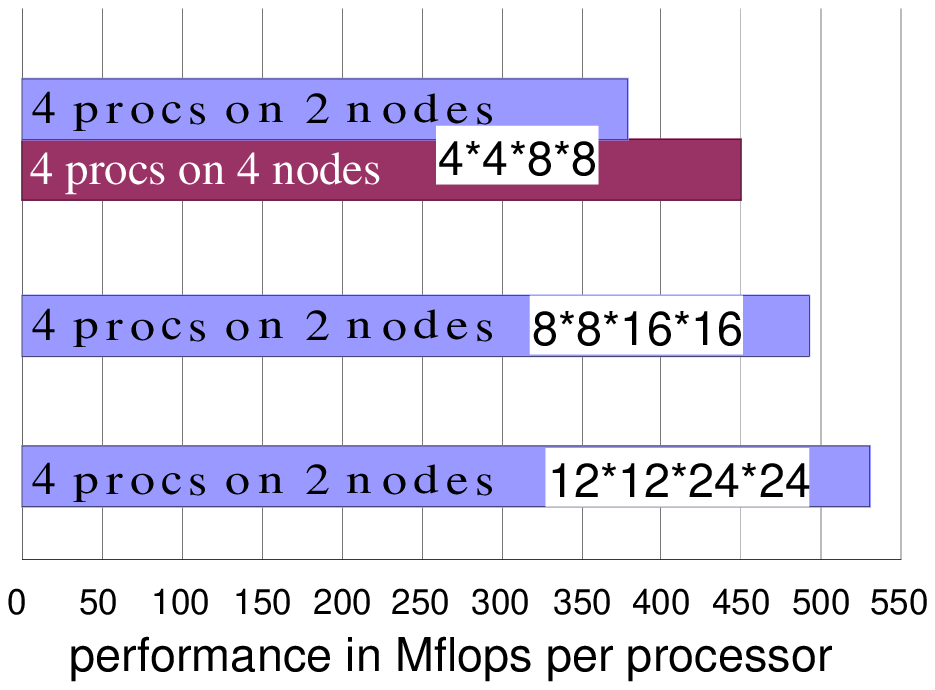}}
\vspace*{-.6cm}
\caption{Parallel performances of double precision 
  e/o Wilson-Dirac matrix-vector multiplications with SSE2 on 4
  processors of the DESY cluster (M.\ L\"uscher, H.\ Wittig)
  \cite{HARTMUTPRIVATE}.}
\label{FIG:HARTMUT}
\end{figure}

Fig.~\ref{FIG:HARTMUT} shows parallel single/dual speeds on the DESY
XEON system, using M.\ L\"uscher's latest version of the e/o
preconditioned Wilson-Dirac matrix-vector multiplication. The
parallelization is 1-dimensional.
With four processors on 2 nodes, a double precision
performance of more than 1 Gflops per node could be achieved.

\begin{figure}[!tb]
\centerline{\includegraphics[width=\columnwidth]{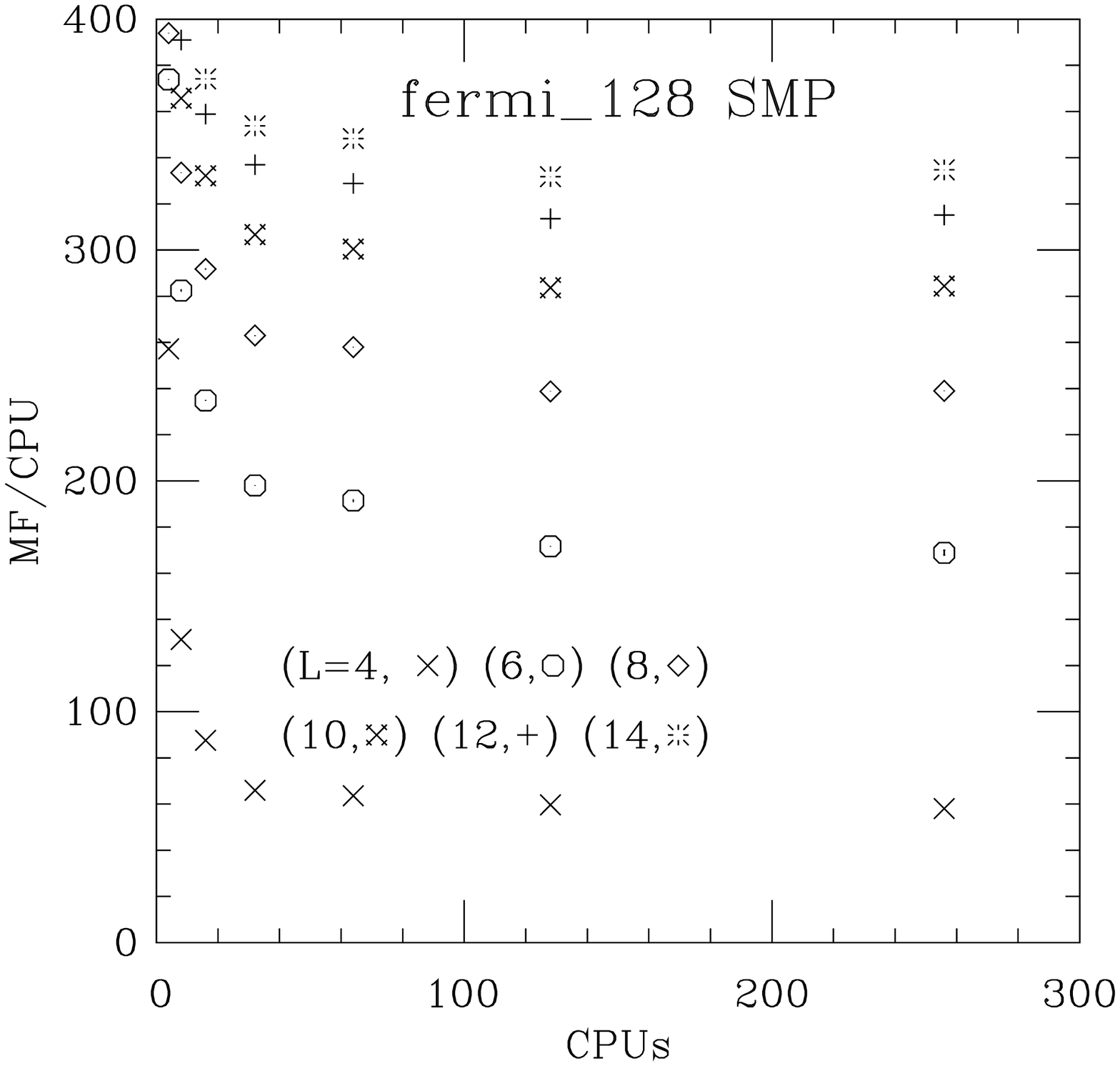}}
\vspace*{-.6cm}
\caption{Efficiency of the MILC staggered fermion code on
  the FNAL dual XEON 128-node cluster.
\cite{GOTTLIEBPRIVATE}.}
\label{FIG:STEVEN}
\end{figure}

Fig.~\ref{FIG:STEVEN} gives an impression of the efficiency of the
MILC staggered fermion code with fixed local lattice sizes on the 128
node dual XEON system at FNAL.

\subsection{Scaling to massive parallelism}
One would like to exert as much CPU power as possible on a given
lattice, as needed, {\em e.g.}, for  realistic turnaround times of dynamical
overlap fermion simulations. While QCDOC and apeNEXT, as shown below, are
designed with respect to fine granularity, clusters favor coarse grained
parallelism.

\begin{figure}[!tb]
\vspace*{-.6cm}
\centerline{\includegraphics[width=\columnwidth]%
{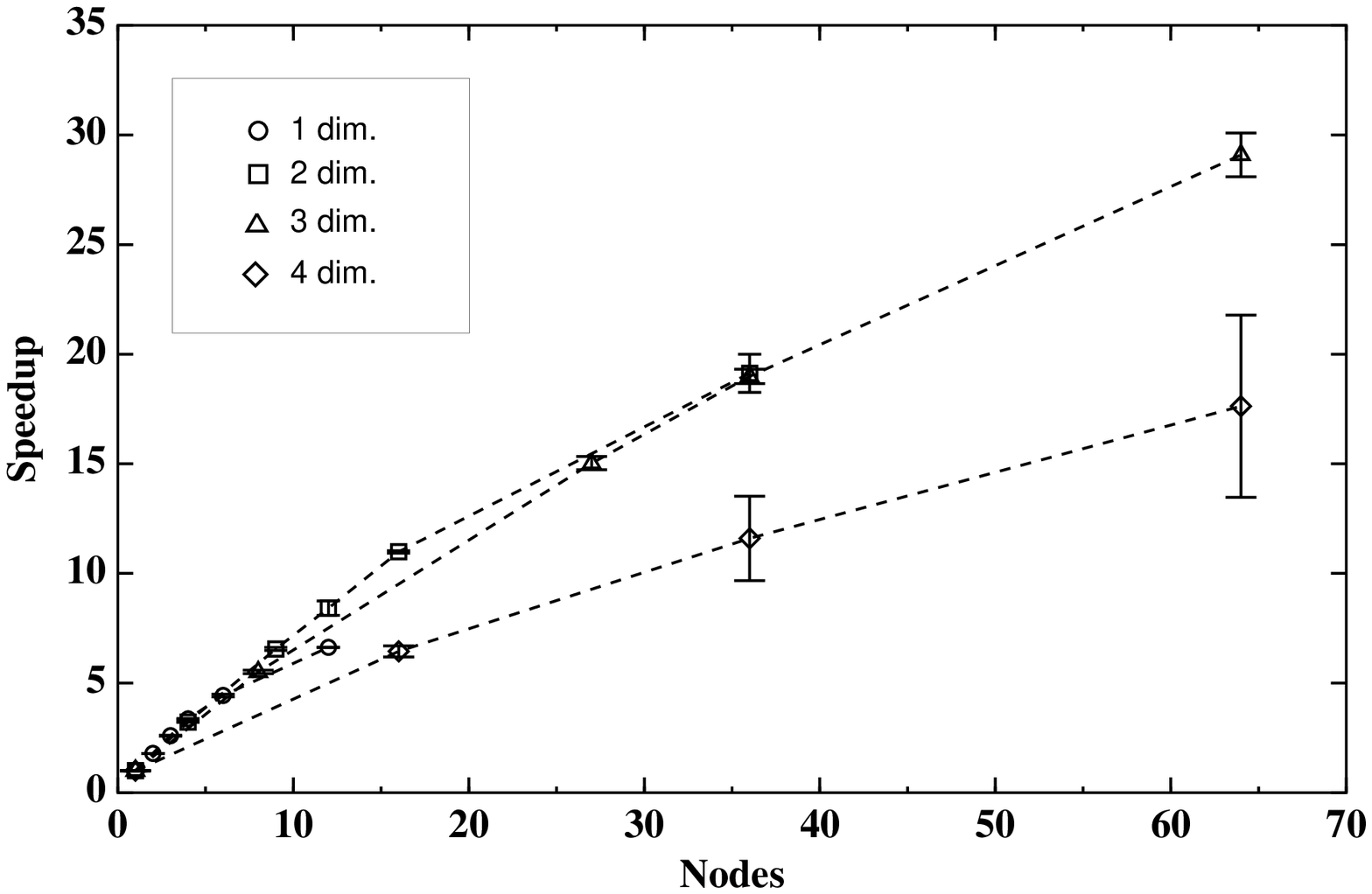}}
\vspace*{-.6cm}
\caption{Scaling of the Wuppertal HMC code with ll-SSOR preconditioned 
  Wilson fermions for a $12^4$ lattice on ALiCE
  \cite{Sroczynski:2002kn,Sroczynski:2003zp}.}
\label{FIG:ZS}
\end{figure}
Nevertheless, Fig.~\ref{FIG:ZS} demonstrates that Wilson fermions with
ll-SSOR preconditioning \cite{Fischer:1996th} and non-blocking MPI
\cite{PARTEC} can scale quite far on clusters. We have benchmarked a
test lattice of size $12^4$.  On 64 ALiCE processors we still achieve
a speedup of about 32 using a 3-d processor geometry.  By
extrapolation we would expect the code to run on 512 processors with a
speedup of 256 for a $16^3\times 32$ lattice.

\section{PROSPECTS\label{APEQCDOC}}

Let me try to assess the current prospects of clusters as compared
to ``home made'' QCD computers. This task is difficult enough as, on one
hand, the development process of home made systems often is
delayed---about two years for APEmille and presumably nearly two years
again both for apeNEXT \cite{Aglietti:1999jp} and QCDOC---, on the
other hand, there are continuous changes in PC processor market.  As
evaluation criteria the cost functions price/performance ratio $R$ for
investment and waste heat $H$ for operation are used.

\subsection{QCDOC}

The development of QCDOC (``QCD on a chip'') is documented in three
proceedings of previous lattice conferences
\cite{Chen:2000bu,Boyle:2001dc,Boyle:2002wg}.  In fact there is very good
news: first successful floating point operations have been carried out on a
prototype ASIC \cite{Boyle:2003ue} in Q2 2003.

The QCDOC CPU is based on a 500 MHz, 32 bit PPC 440 core, with a 64-bit
floating point unit of 1 Gflops peak, and 4 MB on-chip memory.  The
nearest-neighbor topology is a 6-d hypercube with an aggregate bandwidth
of 12 Gbit/s (for 12 directions) per processor. With 550 ns, the latency will be
extremely small. A simulation of the processor gave
a sustained performance of 50\% of peak or 465 Mflops for the
Wilson-Dirac operator on a $2^4$ lattice (T.\ Wettig in
\cite{Hasenbusch:2003rs}).

Due to low latency and high local efficiency, QCDOC will be perfectly
scalable and hence deliver full compute power on small
lattices.  High performances require the use of assembler coding.
Peter Boyle's assembler generator will be an important asset of the
machine.  Large QCDOC systems are likely to be partitioned into
smaller parts.  Main physics targets are dynamical (chiral) fermion
simulations with small quark masses.

At the time of this conference first daughter-boards have been
tested, a 128-node system is planned for autumn 2003, a 5k-node system
should be finished end of 2003. In late spring 2004, 5 Tflops sustained are
planned for both UKQCD/Edinburgh and Riken and 2.5 Tflops sustained for
Columbia University. Funding is aimed at a 10 Tflops sustained QCDOC system
for the US-community (SCIDAC) \cite{NORMANCHRISTPRIVATE}.

\subsection{apeNEXT}

Detailed information on the development of apeNEXT can be found in the
proceedings of earlier lattice conferences
\cite{Alfieri:2001qf,Bodin:2001hn,Ammendola:2002xt}.

The apeNEXT processor design has been finished end of June 2003
\cite{Bodin:2003ww}.  The 200 MHz 64 bit CPU hosts a 64-bit floating
point unit capable of 1.6 Gflops peak.  The memory bandwidth is 3.2
GB/s. The topology is 3-dimensional with an aggregate
nearest-neighbor performance of 1.2 GB/s. The latency will be O(100) ns
and thus favor high scalability.  The processor simulator achieves a
sustained performance of 944 Mflops for the Wilson-Dirac operator (D.\ 
Pleiter in \cite{Hasenbusch:2003rs}). Therefore, 512 processors in a
rack can deliver about 0.5 Tflops sustained. In addition to TAO a C compiler
will be available.  Note that the partitions are quantized 
in units of $4n\times 8\times 8$ processors, $n\in \mathbb{N}$.

A first processor prototype is expected for late 2003. In early 2004,
256 nodes should be assembled.  INFN plans for the funding of several
sustained Tflops, while DESY and GSI (Germany) intend to install 15
and 10 Tflops peak, respectively
\cite{PLEITERPRIVATE,TRIPICCIONEPRIVATE,RAPUANOPRIVATE}.

\subsection{A low-cost QCD-cluster\label{BUILD}}

Let's build our own cost-optimized QCD-cluster!  We adopt the LatFor
setting \cite{Hasenbusch:2003rs} where dynamical Wilson fermions are
simulated by HMC on a $32^3 \times 64$ lattice at a quark mass
characterized through $0.3 < m_{\pi}/m_{\rho} < 0.6$.  To achieve
reasonable turnaround times, we aim for 0.25 Tflops sustained.

Recall that L\"uscher and Wittig got between 380 and 490 Mflops/proc
sustained performance on a dual XEON 2.0 GHz node under Myrinet2000
for local lattice sizes between 1k and 16k sites (section~\ref{PA}).
On a 512-processor system, the local lattice for the LatFor test case
is 4k sites.  Hence it is reasonable to take the average of both
numbers, {\em i.e.}\ 430 Mflops/proc, adding up to a total sustained
performance of 0.22 Tflops.  As connectivity we can choose a
2-dimensional GigE mesh of $32/2 \times 64/2 = 512$ processors since
Fodor et al.\ have demonstrated that GigE meshes come close to Myrinet
performances on the DESY machine \cite{Fodor:2002zi}.

Let us specify the following Gedanken-cluster (prices by
\texttt{www.pricewatch.com}, Q2 2003):
\begin{minipage}{\columnwidth}
\vspace*{.2cm}
\renewcommand{\tabcolsep}{6pt} 
\renewcommand{\arraystretch}{1.1} 
\begin{tabular*}{\columnwidth}{@{}llr}
Mobo &   GA-8EGXR-PEC, 533FSB    &       \\
     &   DDR-266, 6 PCI          &  \$210 \\
CPU &   2 XEON 2.0GHz, 512K CACHE&  \$258 \\
\end{tabular*}
\vspace*{.2cm}
\end{minipage}
\begin{minipage}{\columnwidth}
\vspace*{.2cm}
\renewcommand{\tabcolsep}{14pt} 
\renewcommand{\arraystretch}{1.1} 
\begin{tabular*}{\columnwidth}{@{}llr}
Mem  &   1 GB dual DDR 266 MHz   &  \$119 \\
Case &   incl. Power 500 W       &  \$55  \\
Disk &   EIDE 80 GB              &  \$66  \\
GigE &   4 x PCI cards 4 x \$29  &  \$126 \\
\hline
Sum  &   per dual node           &  \$834 \\
\hline
\end{tabular*}
\vspace*{.2cm}
\end{minipage}
The waste heat, $H$, amounts to about 30 kW.

\subsection{Comparison}

Table.~\ref{TAB:COMPARE} confronts cost functions and maximal
processor numbers of QCDOC, apeNEXT and mesh cluster with respect
to the LatFor test case.
\begin{table}[!b]
\vspace*{-.3cm}
\caption{Cost functions for  QCDOC, apeNEXT and 
Cluster with respect to the LatFor test case.
$P$ is the total
performance in Tflops, $R$ is the price/performance ratio in \$/Mflops, $H$
the waste heat in W/Mflops, and $C$ the maximal number of processors.
Performances are sustained.}
\label{TAB:COMPARE}
\renewcommand{\tabcolsep}{3pt} 
\renewcommand{\arraystretch}{1.1} 
\begin{tabular*}{\columnwidth}{@{}lllllcr}
\hline
system      & year & proc & $P$ & $R$ & $H$  & $C$ \\
\hline
QCDOC       & 2004   & 512 & 0.238 & $\approx$1 & 0.01  & $>16k$   \\
apeNEXT     & 2004/5 & 256 & 0.241 & $\approx$1 & 0.02  & $2048$   \\
Cluster     & 2003   & 512 & 0.220 & $\approx$1 & 0.12  & $512$    \\
\hline
\end{tabular*}
\end{table}

\noindent\fbox{$R$} First we  extrapolate $R$ to equal points in time,
  say 01/2005.  $R$ is likely to drop to 0.5 \$/Mflops for the cluster
  system by then (Moore's ``Law'').  Hence, investment costs will
  favor a cluster in 01/2005.
  
\noindent\fbox{$H$} The cost of operation of the cluster as determined through
  $H$ will lie below \$20.000 per year, assuming German electricity
  costs for major customers. Operating QCDOC and apeNEXT will be
  considerably cheaper by a factor of 10 and 5, respectively. Thus, costs of
  operation favor QCDOC or apeNEXT. 
  
\noindent\fbox{$C$} With respect to the LatFor test case, the maximal
number of processors, $C=512$, that can be realized  for a 2-d mesh geometry has
been chosen. In contrast, apeNEXT is limited to $C=2048$ processors 
while QCDOC can deploy tens of thousands of processors.

In order to improve on this situation for clusters, one can resort to
a 3-dimensional geometry, which in principle allows for $C=8k$. Of
course, the scalability $S$ might limit the performance for yet
smaller numbers of processors. It is possible that cache-resident
coding will help here  \cite{WATSONPRIVATE}.

As far as dynamical Overlap fermions are concerned, first simulations
are likely to use lattices of size $16^3\times 32$.  Aiming at maximal
throughput, one should be aware that the numbers of processors are limited to
$C=128$ for a 2-d cluster, $C=1$k for a 3-d cluster, $C=1$k for
apeNEXT, or $C=8$k for QCDOC. In other words, the 8k QCDOC can
simulate this specific Overlap fermion problem 4 times as fast as the
2-crate apeNEXT, 8 times as fast as the 3-d cluster with 1k processors
(assuming scalability) or 64 times as fast as the 2-d cluster with 128
processors.

\section{SUMMARY AND OUTLOOK}

The price/performance ratio of QCD-clusters has just crossed the
$R=1$\$/Mflops threshold, QCDOC and apeNEXT are supposed to deliver
this ratio mid/end of 2004. The waste heat per Mflops, $H$, is about
10 times larger for clusters.  Hence, the TCO for 5 years of operation
turns out to be similar for QCDOC, apeNEXT and clusters.

As far as simulations on small lattices are concerned, the attainable
throughput depends on the compute power applicable which is determined
by the dimensionality of the parallelization. This is an advantage of
3-d and 4-d network geometries.

Clusters can be used for complicated actions if a switched network
complements mesh or grid. They will further improve with respect to
home made systems due to PCI-Express\footnote{PCI-Express based
  co-processors like the 25 Gflops ClearSpeed\texttrademark\ CPU-array
  just announced \cite{CLEARSPEED} might be promising PC accelerators.},
networks and improved communication software.

Jefferson Lab has installed a GigE-mesh QCD-cluster these days with
256 dual XEON nodes arranged as a $4\times 8\times 8$ grid, expected
to deliver 1 Gflops per node sustained for the Wilson-Dirac operator
\cite{WATSONPRIVATE}.  Wuppertal University is about to install a 1024
processor system combining a GigE mesh architecture with a switched
network.

At last, we should gauge all our efforts with respect to commercial
supercomputers scheduled for 2\$/Mflops sustained end of 2005.

\subsection*{Acknowledgments} 
I thank the organizers of Lattice 2003 for their kind invitation.  I
am indebted to Norbert Eicker and Klaus Schilling for important
discussions and careful reading of the manuscript.

\end{document}